\documentclass[A4,12pt]{article}
\usepackage{graphicx}
\begin{document}

\title{Relativistic Solutions of  Anisotropic Compact Objects}

\author{Bikash Chandra Paul$^{a,c}$\thanks{e-mail : bcpaul@iucaa.ernet.in},    Rumi Deb$^{b}$ \\
 $^a$Physics Department, North Bengal University, \\
Siliguri, Dist. : Darjeeling, Pin : 734 013, West Bengal, India
\\
$^b$IUCAA Resource Centre, Physics Department \\
PO : North Bengal University, Dist.: Darjeeling, Pin : 734013, India \\
$^c$ Institute of Theoretical Physics, KITPC \\
Chinese Academy of Sciences, Beijing 100190, China}

\maketitle
\date{}

\begin{abstract}
We present a class of new relativistic  solutions with anisotropic fluid for  compact stars  in hydrostatic equilibrium. The interior space-time geometry considered here for  compact objects are described by parameters namely,  $\lambda$, $k$, $A$, $R$ and $n$. The values of the  geometrical parameters are determined here for obtaining a class of physically viable stellar models. The energy-density, radial pressure and tangential pressure are finite and positive inside the anisotropic stars. Considering some stars of  known mass we present stellar models which  describe compact astrophysical objects with nuclear density.
\vspace{0.1 cm}

PACS No(s). 04.20.Jb, 04.40.Dg, 95.30.Sf
\end{abstract}


\section{Introduction:}

The precision astronomical observations in the  last couple of decades predicted the existence of massive compact objects.  A number of compact objects with very high densities are discovered in the recent times \cite{tab}. To describe such compact objects general theory of relativity is most useful. The theoretical investigation of such compact  astrophysical objects has been a key issue in relativistic astrophysics over a couple of decades. Astrophysical objects with perfect fluid necessarily requires the  pressure inside  is  isotropic \cite{bv}. In general, a polytropic equation of state (EOS) is   used widely to describe a white dwarf or a less compact star   \cite{sl}. However, theoretical understanding in the last couple of decades made it clear that there is a deviation from local isotropy in the interior pressure. At very high enough densities with smaller radial size the anisotropic pressure plays an important role in determining stellar properties \cite{rr,vc,ltm}. The physical situations where anisotropic pressure may be relevant are very diverse for a compact stellar object \cite{rr,vc,mm,bm}. By anisotropic pressure we mean  the radial component of the pressure ($p_r$) different from that of the tangential pressure $p_t$.
 	After the seminal  work of Bowers and Liang \cite{bm}, a number of literature appeared considering an   anisotropic spherically symmetric static general relativistic object.   \cite{rr} and  \cite{vc} theoretically investigated   compact objects and observed that a star with matter  density   ($\rho > 10^{15} gm/cc$), where the nuclear interaction become relativistic in nature, are likely to be anisotropic.
It is further noted that anisotropy in fluid pressure in a star  may originate  due to  number of processes  e.g., the existence of a solid core, the presence of type 3A super fluid etc. \cite{kw}.
  Recently, \cite{mh} determined the maximum mass and mass to radius ratio of  a compact isotropic relativistic star. \cite{mm}, \cite{bm}, \cite{sb} examined  spherical distribution of anisotropic matter in the framework of general relativity and derived a number of solutions to understand the interior  of such stars. A handful  number of  exact interior solutions in general relativity  for both the isotropic and the anisotropic compact objects have been reported  in the literature \cite{dl}.   \cite{dl} analysed  127 published solutions out of which they found that only 16 of the published results  satisfy all the conditions for a physically viable stellar  model. In the case  of a  compact stellar object it is essential to  satisfy all the conditions outlined by Delgaty and Lake as the EOS of the fluid of the compact dense object is not known.

The discovery of compact stellar objects, such as X-ray pulsars, namely Her X1, millisecond pulsar SAX J1808.43658, X-ray sources,  4U 1820-30 and 4U 1728-34 are important and interesting as  these are considered to be probable strange star candidates.  The existence of such characteristics compact objects  led to critical studies of  stellar configurations \cite{dd,li,kt,ml,mpd,nd,bombaci,th,rt,gj,jt,tik,fs}. However, the equation of state (EOS) of matter inside a superdense strange star at present is not known. In this context \cite{vt} and  \cite{rt} have shown that in the absence of definite information about the EOS of matter content of stellar configuration, an alternative approach of prescribing a suitable {\it ansatz} for the  geometry of the interior physical 3-space of the configuration leads to simple easily accessible physically viable  models of such stars.  Relativistic models of superdense stars based on different solutions of Einstein's field equations obtained by Vaidya-Tikekar approach of assigning different geometries with physical 3-spaces of such objects are reported in the literature \cite{kt,mpd,th,jt,tik}. \cite{ps} obtained a class of relativistic static non-singular analytic solutions in isotropic form with a spherically symmetric  distribution of matter  in a  static  space time.  Pant and Sah  solution is found to lead to a  physically viable causal model of neutron star  with a maximum mass  of $4 M_{\odot}$. Recently,  \cite{rb} obtained a class of compact stellar models using Pant and Sah solution in the case of spherically symmetric space time. In this paper we obtain a class of new relativistic solutions which accommodate  anisotropic stars possessing mass relevant for neutron stars. Usually a stellar model is obtained using Einstein field equation for a known EOS and then the geometry of the space-time is determined. In this paper we  follow an alternative approach (Synge approach) by first making an {\it ad hoc} choice of the geometry and then explore the EOS for matter. A class of new relativistic solutions are discussed here which accommodate anisotropic star in hydrostatic equilibrium having mass and radius relevant for neutron stars \cite{slb}.

The paper is organised as follows: In section 2, we set up the relevant field equations and its solutions. In section 3, physical properties of anisotropic star is presented. In sec. 4, we present  physical analysis of stellar models with the observational stellar mass  for different model parameters. Finally  in sec 5, we give a brief discussion.

\section{Field Equation and Solutions}

The Einstein's field equation is
\begin{equation}
R_{\mu \nu} -\frac{1}{2} g_{\mu \nu} R =  8 \pi G  \; T_{\mu \nu}
\end{equation}
where  $g_{\mu \nu}$, $R$, $R_{\mu \nu}$ and $T_{\mu \nu}$ are the  metric tensor, Ricci scalar,  Ricci tensor and energy
momentum tensor respectively.

We use spherically symmetric  space time metric given by
\begin{equation}
ds^2= e^{\nu(r)}dr^2-e^{\mu(r)}(dr^2+r^2d\Omega^2)
\end{equation}
where $\nu(r)$ and $\mu(r)$ are  unknown metric functions and  $d\Omega^2=d\theta^2+ sin^2\theta \; d\phi^2 $. We assume an anisotropic pressure distribution for the fluid content of the star. The energy momentum tensor for such  fluid  in equilibrium is given by
\begin{equation}
T^{\mu}_{\mu} = diag \;(\rho, - p_r,- p_t, - p_t)
\end{equation}
 where  $\rho$ is the energy-density, $p_r$ is the radial pressure, $p_t$ is the tangential pressure and $\Delta=p_t-p_r$ is the measure of pressure anisotropy \cite{slb}. Using the space time metric given by eq.(2), the Einstein's field eq. (1) reduces to  the following equations:
\begin{equation}
\rho=-e^{-\mu}\left(\mu''+\frac{\mu'^2}{4}+\frac{2\mu'}{r}\right),
\end{equation}
\begin{equation}
p_r=e^{-\mu}\left(\frac{\mu'^2}{4}+\frac{\mu'}{r}+\frac{\mu'\nu'}{2}+\frac{\nu'}{r}\right),
\end{equation}
\begin{equation}
p_t=e^{-\mu}\left(\frac{\mu''}{2}+\frac{\nu''}{2}+\frac{\nu'^2}{4}+\frac{\mu'}{2r}+\frac{\nu'}{2r}\right).
\end{equation}
Using eqs. (5) and (6) along with the definition of anisotropy of fluid we obtain
\begin{equation}
\left(\frac{\mu''}{2}+\frac{\nu''}{2}+\frac{\nu'^2}{4}-\frac{\mu'^2}{4}-\frac{\mu'}{2r}-\frac{\nu'}{2r}-\frac{\mu'\nu'}{2}\right)=\Delta e^{\mu}.
\end{equation}
Eq. (7) is a second-order differential equation which admits a class of new solution with anisotropic fluid distribution given by
\begin{equation}
e^{\frac{\nu}{2}}=A \left(\frac{1-k\alpha+n\frac{r^2}{R^2}}{1+k\alpha } \right), \;\;\; \;\;\;  e^{\frac{\mu}{2}}=\frac{(1+k\alpha)^2}{1+\frac{r^2}{R^2}}
\end{equation}
where
\begin{equation}
\alpha(r)=\sqrt{\frac{1+\frac{r^2}{R^2}}{1+ \lambda \frac{ r^2}{R^2}}}
\end{equation}
with $R$, $\lambda$, $k$, $A$ and $n$  are  arbitrary constants.  It may be pointed out here that $n=0$ corresponds to a solution for isotropic stellar model obtained by \cite{ps}. We consider here non-zero $n$ to obtain an anisotropic stellar model in hydrostatic equilibrium. Eq.(8) permits a relation amongst the parameters which is useful for obtaining stellar models. The allowed values of the parameters are determined using the physical conditions imposed on the stellar solution for a viable stellar model.
The geometry of  the 3-space in the above  metric is given by
\begin{equation}
d\sigma^2=\frac{dr^2+r^2(d\theta^2+sin^2\theta d\phi^2)}{1+\frac{r^2}{R^2}}.
\end{equation}
It corresponds to  a 3 sphere immersed in a 4-dimensional Euclidean space. Accordingly the geometry of physical space obtained at the $t=constant$  section of the space time is given by
\[
ds^2=A^2\left(\frac{1-k\alpha+n\frac{r^2}{R^2}}{1+k\alpha}\right)^2 dt^2
\]
\begin{equation}
\; \; \; \;  -\frac{(1+k\alpha)^4}{(1+\frac{r^2}{R^2})^2} \left[dr^2+r^2(d\theta^2+sin^2\theta d\phi^2)\right].
\end{equation}
The pressure anisotropy term becomes
\begin{equation}
\Delta=  \frac{2 n \frac{r^2}{R^2} (8 \alpha (1+ \lambda \frac{r^2}{R^2} )^3 + k^2 \alpha X  + Y  )  }{ \alpha^{3/2} (1+\lambda \frac{r^2}{R^2})^4 (1+k \alpha)^2 (1 + n \frac{r^2}{R^2} - k \alpha) }
\end{equation}
where
$X=  8 \lambda^2 \frac{r^6}{R^6} + 4 \lambda (1+5 \lambda) \frac{r^4}{R^4}+12 \lambda -4,
\; \;
Y=(15 \lambda^2+10 \lambda -1) \frac{r^2}{R^2} +k(4+12 \lambda+16 \lambda^2) \frac{r^6}{R^6} + 4 \lambda (5+ 7 \lambda ) \frac{r^4}{R^4} +(15 \lambda^2+26 \lambda +7)\frac{r^2}{R^2} $.
The geometry of 3 - space obtained at $t=constant$ section of the space time  metric (11) given above incorporates a  deviation  in a spherical 3 space,  $k$ is a geometrical parameter measuring inhomogeneity of the physical space and $n$ is related to the anisotropy.  For $k=0$ and $n=0$, the space time metric (11) degenerates into that of Einstein's static universe which is filled with matter of uniform density. The solution obtained by Pant and Sah corresponds to the case when $n=0$ and $k\neq 0$ \cite{ps}. It reduces to a generalization of the Buchdahl solution, the physical 3-space associated with which has the same feature.
 However, for $\lambda>0$, the solution corresponds to finite boundary models. In this paper we study physical properties of compact objects filled with anisotropic fluid ($n\neq 0$) and determine the  values of   $R$, $\lambda$, $k$ and  $A$ for a viable stellar model as permitted by the field equation.
The exterior Schwarzschild line element is given by
\[
ds^2= \left( 1 - \frac{2m}{r_o} \right) dt^2 - \left( 1 - \frac{2m}{r_o} \right)^{-1} dr^2
\]
\begin{equation}
\; \; \; \; \; \; - r_o^2 (d\theta^2 + sin^2 \theta d\phi^2)
\end{equation}
where $m$ represents the mass of spherical object. The above metric can be expressed in an isotropic metric form \cite{jvn}
\begin{equation}
ds^2=\left(\frac{1-\frac{m}{2r}}{1+\frac{m}{2r}}\right)^2dt^2-\left(1+\frac{m}{2r}\right)^4(dr^2+r^2d\Omega^2)
\end{equation}
using the transformation $r_o= r \left(1+ \frac{m}{2r} \right)^2$  where $r_o$ is the radius of the compact object. This form of the Schwarzschild metric will be used here to match at the boundary with the interior metric given by eq. (11) at the boundary.

\section{Physical properties of  anisotropic compact star}

The solution given by eq.(8) is useful to study physical features of  compact objects with anisotropy in a general way which are outlined as follows:

(1) In this model, a positive central density $\rho$ is obtained  for $\lambda<\frac{4}{k}+1$.

(2) At the boundary of the star ($r=b$), the interior solution should be matched with the isotropic form of Schwarzschild exterior solution, {\it i.e.},
\begin{equation}
e^{\frac{\nu}{2}}|_{r=b}=\left( \frac{1-\frac{m}{2b}}{1+\frac{m}{2b}}\right)\;\;;\;  e^{\frac{\mu}{2}}|_{r=b}=\left(1+\frac{m}{2b}\right)^2
\end{equation}

(3) The physical radius of a star ($r_o$), is determined  knowing the radial distance where  the pressure at the boundary vanishes (i.e., $p(r)=0$ at $r=b$). The  physical radius is related to the radial distance ($r=b$) through the relation $r_o= b \left(1+ \frac{m}{2b} \right)^2$ \cite{jvn}.

(4) The ratio $\frac{m}{b}$ is determined using eqs. (8) and (14), which is given by
\begin{equation}
\frac{m}{b} = 2 \pm 2A\left( \frac{1-k\alpha+n y^2}{\sqrt{1+y^2}} \right)
\end{equation}
where $
y=\frac{b}{R}$.
In the above we consider only negative sign as it corresponds  to a physically viable stellar model.

(5) The density inside the star should be positive i.e., $\rho>0$.

(6) Inside the star the stellar model should satisfy the condition, $\frac{dp}{d\rho}<1$ for the sound propagation to be causal.

The usual boundary conditions are that the first and second fundamental forms required to be continuous across the boundary $r=b$. We determine $n$, $k$, $\lambda$ and $A$ which satisfy the criteria for a viable stellar model outlined above.
As the field equations are highly non-linear and intractable to obtain a known functional relation  between pressure and density we adopt numerical technique.
Imposing the condition that the  pressure at the boundary vanishes, we determine $y$  from  eq. (5).
The square of the acoustic speed  $\frac{dp}{d\rho}$ becomes :
\begin{equation}
\frac{dp}{d\rho}=-\frac{\sqrt{\alpha}(1+k\sqrt{\alpha})(A+\frac{B}{\sqrt{\alpha}}+C+D)}{E}
\end{equation}
where
\[A = -4 (-1+n+n^2+2n^2r^2+nr^4)(1+r^2\lambda)^5
\]
$+2k^4(1+r^2)^4\lambda(-1+3(3+2r^2)\lambda)$
\[
B=2k^3(1+r^2)^3((1+r^2)(\lambda-1)\lambda+
\]
$n(1+(1+4r^2+2r^4)\lambda-r^4\lambda^2
+r^4(3+4r^2+2r^4)^3\lambda)),
$
\[
C=k\sqrt{\alpha}(1+r^2\lambda)^3(-2(-1+\lambda-r^2\lambda+r^2\lambda^2)
\]
$-n(-6+10\lambda+8r^6\lambda+4r^8\lambda^2+r^2(-21
+34\lambda-5\lambda^2)+r^4(-5+12\lambda+\lambda^2))+n^2(-8+4r^8(\lambda-1)\lambda-2r^2(9+7\lambda)+r^6(5-26
\lambda+5\lambda^2+r^4(3-52\lambda+9\lambda^2)))),
$
\[D=k^2(1+r^2)(1+r^2\lambda)(-2(5+(4r^2-7-4r^4)\lambda
\]
$+(6+8r^2+19r^4+2r^6)\lambda^2
+r^4(-3+2r^2)\lambda^3)+n^2(1+r^2)(4r^8(\lambda-1)\lambda^2-4-2r^2(3+5\lambda)
+r^6\lambda(3\lambda^2-5-6\lambda)-r^4(1+16\lambda+3\lambda^2))+n(12-8\lambda+12r^8\lambda^3+4r^10\lambda^3+r^2
(42\lambda-1-29\lambda^2)+r^6\lambda(5+2\lambda+9\lambda^2)+r^4(25\lambda-3+3\lambda^2-9\lambda^3))),
$
\[
E=6(1+nr^2-k\sqrt{\alpha})^2(2\sqrt{\alpha}(1+r^2\lambda)^5
\]
$+k^3(1+r^2)^4\lambda(-1+(3+2r^2)\lambda)+2k^2(1+r^2)\sqrt{\alpha}(2+(3r^2-3-2r^4)\lambda+(4+5r^2+13r^4)\lambda^2+r^2(4+7r^2+13r^4+2r^6)\lambda^3+r^6(r^2-1)\lambda^4)+k(1+r^2)(6+(-5+16r^2-3r^4)\lambda+(5+3r^2+33r^4-r^6)\lambda^2+r^2(5+6r^2+27r^4+2r^6)\lambda^3+(4r^8-2r^6)\lambda^4)).
$

We study the physical properties of anisotropic compact objects numerically and follow the following steps. For given values of $\lambda$ and $k$, the size of the star is estimated from the condition that pressure vanishes at the boundary which follows from eq.(5). The mass to radius  $\frac{m}{b}$ of a star is determined from eqs.(8) and (14), which in turn determines the physical size of the compact star ($r_o$).  For  a given set of values of the parameters $\lambda$, $A$, $k$,  $n$, and the mass ($m$), the radius of  an anisotropic compact object is obtained  in terms of the model parameter $R$. Thus for a known mass of a compact star $R$ is determined which in turn determines the corresponding radius.
\\

\begin{figure}
\begin{center}
\includegraphics{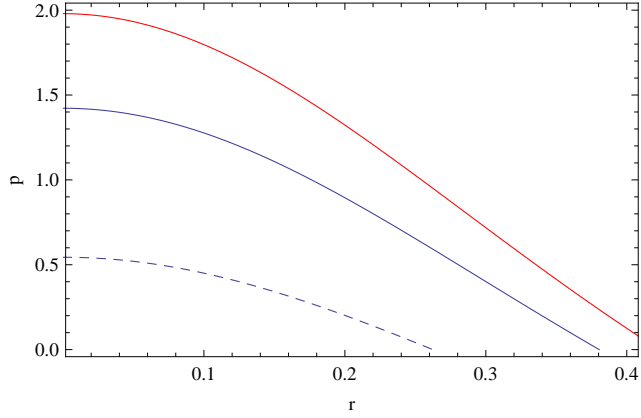}
\caption{Radial variation of pressure for different $k$ with $n=0.60$, $\lambda=1.9999$ and $A=2$. Red  line for $k = 0.55$, blue line for $k = 0.5$ and dashed  line for $k=0.4$.}
\end{center}
\end{figure}

\begin{figure}
\begin{center}
\includegraphics{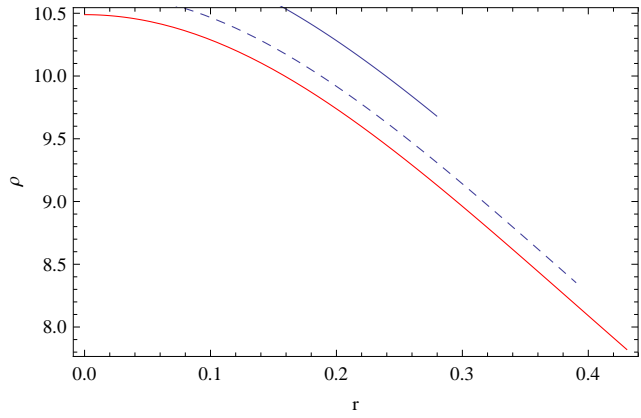}
\caption{Radial variation of density for different $k$ with $n=0.60$, $\lambda=2$ and $A=2$.  Blue  line for $k = 0.40$, dashed line for $k = 0.50$ and red line for $k=0.55$.}
\end{center}
\end{figure}

\begin{figure}
\begin{center}
\includegraphics{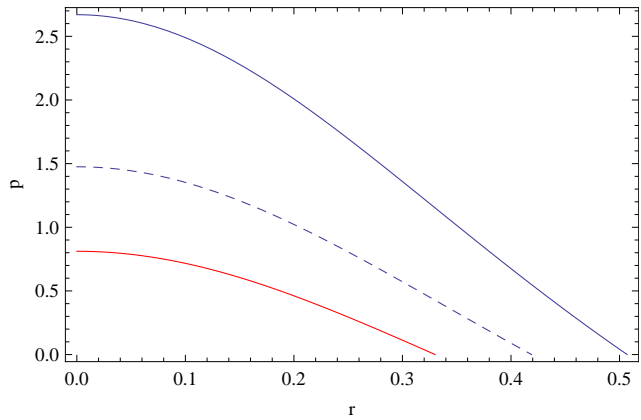}
\caption{Radial variation of pressure for different $n$ with $k=0.31$, $\lambda=2$ and $A=2$.  Blue  line for $n = 1.22$, dashed line for $n = 0.95$ and red line for $n=0.8$.}
\end{center}
\end{figure}

\begin{figure}
\begin{center}
\includegraphics{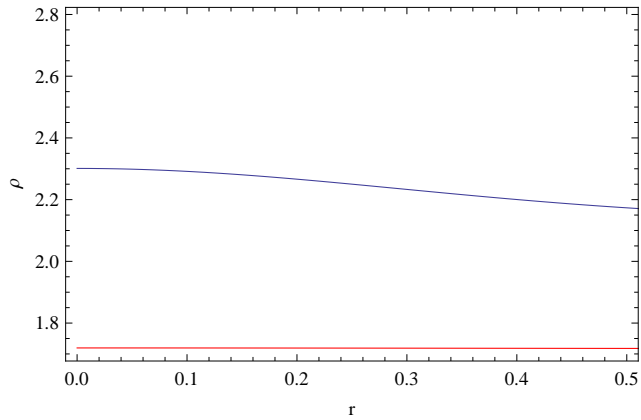}
\caption{Radial variation of density for different $\lambda$ with $k=0.641$, $n=0.60$ and $A=2$.  Blue  line for $\lambda =1.9999 $,  and red line for $\lambda=1.1$.}
\end{center}
\end{figure}

\begin{figure}
\begin{center}
\includegraphics{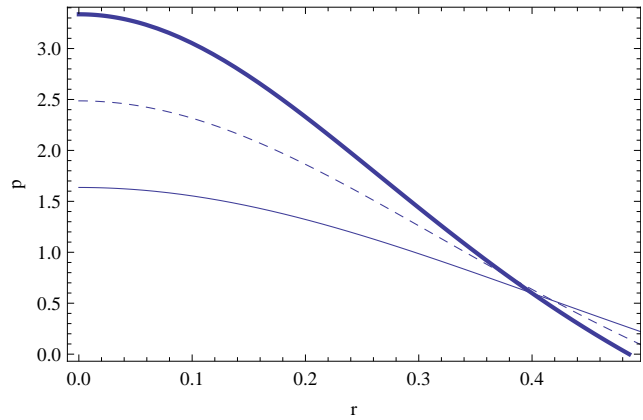}
\caption{Variation of radial pressure for different $\lambda$ with $k=0.641$, $n=0.60$ and $A=2$.  Blue  line for $\lambda = 1.0$, dashed line for $\lambda = 1.5$ and thick line for $\lambda=1.9999$.}
\end{center}
\end{figure}

\begin{figure}
\begin{center}
\includegraphics{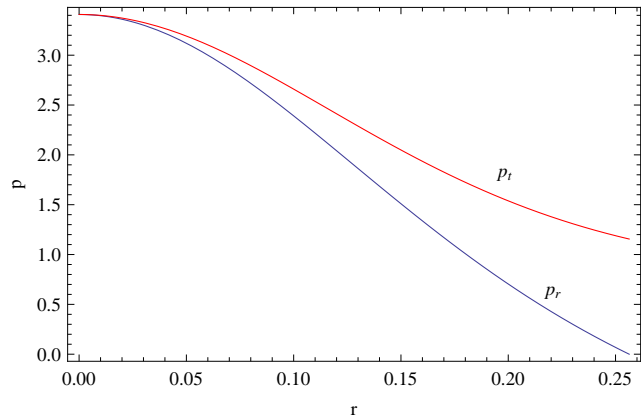}
\caption{Radial variation of transverse and radial pressure  with $\lambda=10$, $n=0.8$, $A=2$ and $k=0.31$.  Blue line for radial pressure and red line for transverse pressure.}
\end{center}
\end{figure}

\begin{figure}
\begin{center}
\includegraphics{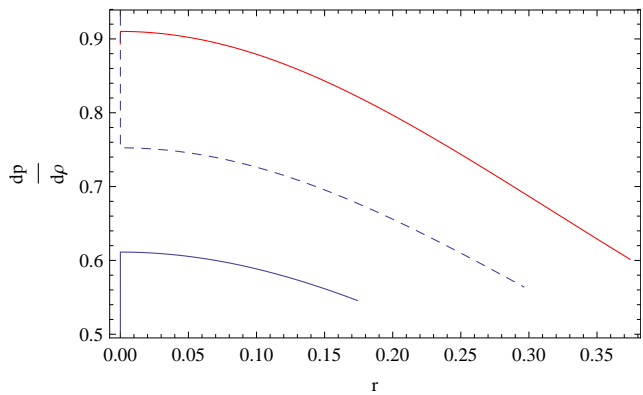}
\caption{Radial variation of $\frac{dp}{d\rho}$ with different $n$ for $k=0.61669$, $\lambda=2$, $A=2$. Red line for $n=0.4$ , dashed line for $n=0.3$ and Blue line for $n=0.2$.}
\end{center}
\end{figure}

\begin{figure}
\begin{center}
\includegraphics{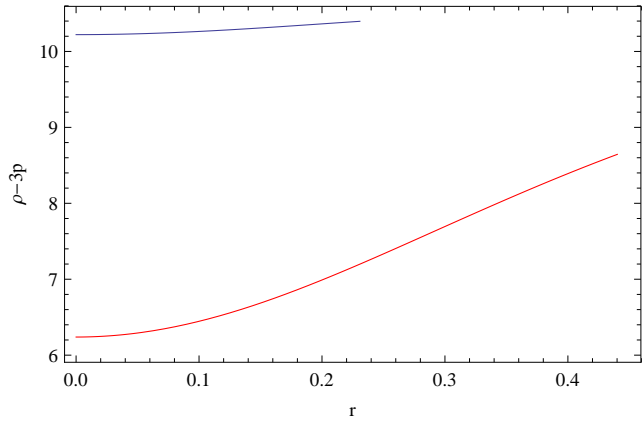}
\caption{Variations of parameter $n$ with radial distant $r$ (in km.) for  SEC $(\rho-3p)$. Blue line for $n=0.7$ and red line for $n=1$.}
\end{center}
\end{figure}

\begin{figure}
\begin{center}
\includegraphics{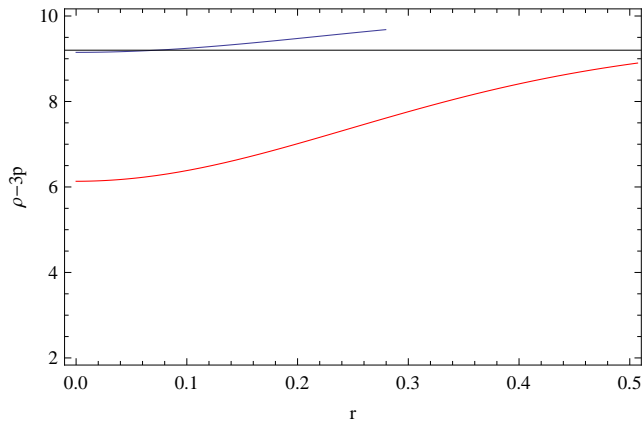}
\caption{Variations of parameter $k$ with radial distance $r$ (in km.) for  SEC $(\rho-3p)$. Blue line for $k=0.4$ and red line for $k=0.50$.}
\end{center}
\end{figure}

\begin{figure}
\begin{center}
\includegraphics{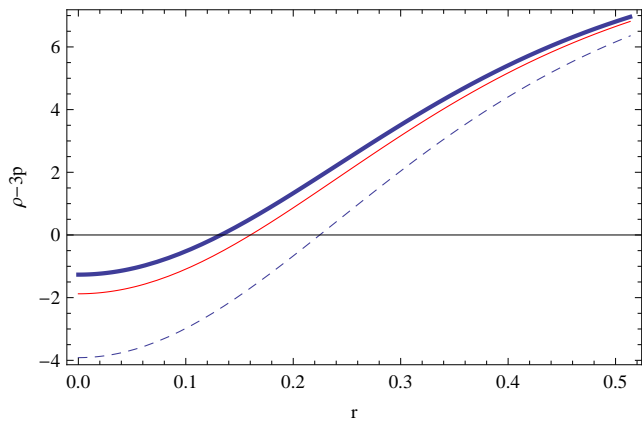}
\caption{Radial variations of SEC {\it i.e.}, $(\rho-3p)$ with different $n$  for $k=0.641$, $\lambda=2$ and  $A=2$.  Dashed  line for $n=0.8$, red line for $n=0.7$ and thick line for $n=0.67$ .}
\end{center}
\end{figure}

\begin{figure}
\begin{center}
\includegraphics{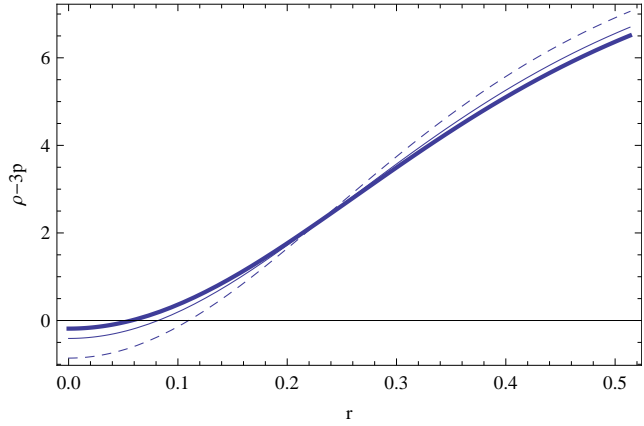}
\caption{Radial variations of SEC  {\it i.e.}, $(\rho-3p)$ with different $\lambda$  for $k=0.641$, $n=0.65$ and  $A=5$.  Dashed  line for $\lambda=2$, blue line for $\lambda=1.8$ and thick line for $\lambda=1.7$ .}
\end{center}
\end{figure}

\begin{figure}
\begin{center}
\includegraphics{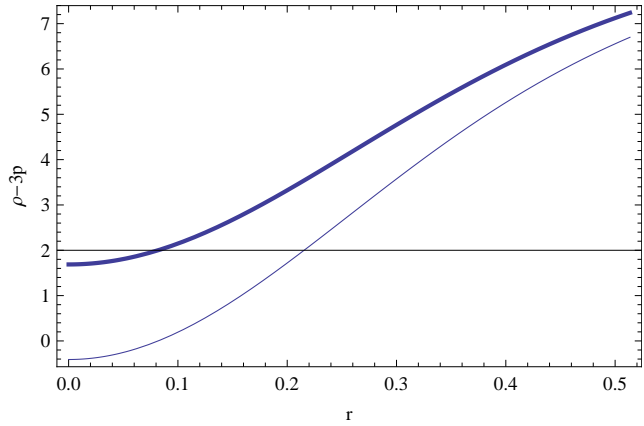}
\caption{Radial variations of SEC {\it i.e.}, $(\rho-3p)$ with different $k$  for $n=0.65$, $\lambda=1.8$ and  $A=5$.  Blue line for $k=0.641$ and thick line for $k=0.6$.}
\end{center}
\end{figure}

\begin{figure}
\begin{center}
\includegraphics{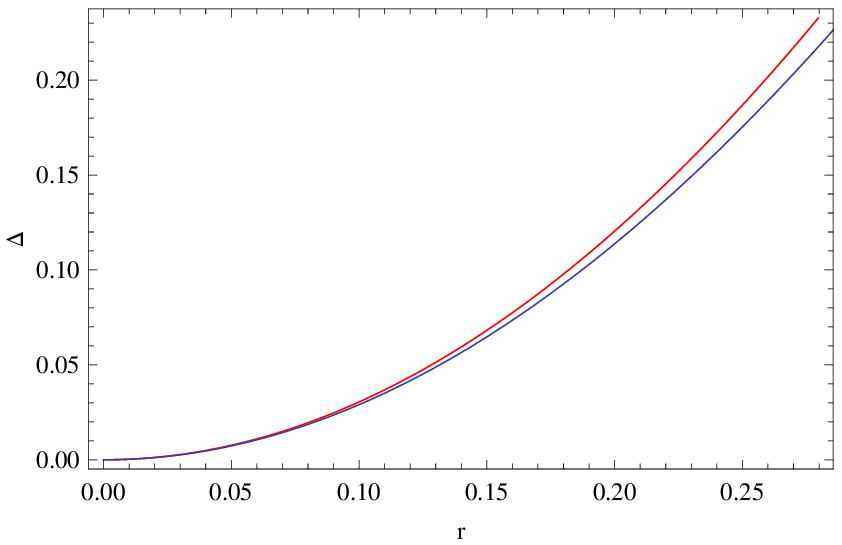}
\caption{Radial variations of anisotropic  parameter $\Delta$ for  different $n$. Blue line for $n=0.7$ and red line for $n=1$.}
\end{center}
\end{figure}

The  radial variation of  pressure and  density of anisotropic  compact objects  for  different parameters  are plotted in figs. (1)-(4). In figs. (1) and (2),  variation  of radial  pressure is  plotted   for a given set of values of $A$, $n$ and $\lambda$ for different  $k$.  It is noted that the  pressure  increases with an increase in $k$ whereas the density decreases. The central density also found to increases with   decrease in the value of $k$. The radial variation of pressure with $n$ is plotted in fig.(3). It is evident that  although the pressure inside the star decreases with an increase in  $n$,  the density remains invariant.  The radial variation of density with $\lambda$ is plotted in fig. (4). Both the density and the pressure are  found to increase with an increase in $\lambda$ value showing an increase in corresponding central density. But the difference between central density with that of surface density reduces with increase in $\lambda$.
 It is noted that both the  pressure and the density are independent on $A$.  The radial variation of pressure for different $\lambda$ is shown in fig. (5), it is evident that the decrease in radial pressure near the boundary is sharp  for higher values of $\lambda$. The variation of both radial and transverse pressure are plotted in fig. (6), it is noted that the value of transverse  pressure at the boundary  is  more than that of radial pressure although they begin with same central pressure at the centre.  Fig. (7) is a plot of squared speed of sound {\it i.e.,}  $\frac{dp}{d\rho}$ with different $n$ values. It is found that $\frac{dp}{d\rho}$ is positive inside the star and obeys causality condition. It shows stability of the stellar models. To check the strong energy condition we plot the radial variation of $(\rho-3p)$  for different values of $n$, $\lambda$ and $k$ values in figs. (8)-(12). In  Figs. (8) and (9)  it is observed the SEC is obeyed. But from  figs. (10)-(12), it is noted  that there exist a region near the center of the star where SEC is not obeyed. It is further noted that the radius of that region   increases with an increase in the parameter values  $n$, $k$ and $\lambda$.  This is interesting as two distinct regions are found to exist in the compact objects corresponding to the solution obtained here which may be useful for constructing a core-envelope model of the star. The  radial variation of anisotropy inside the star for different $n$ values are plotted in fig. (13). It is evident that the  anisotropy of a star increases with increase in value of the parameter $n$.

\begin{table}
\begin{center}
\begin{tabular}{|c|c|c|c|c|}   \hline
$\lambda$               & $n=0 $	&$n=0.55$	&$n=0.58$	&$n=0.60$     \\ \hline
4   &  0   		& 0.333416		& 0.342962		& 0.34913	\\ \hline
4.1 &  0.051703	 & 0.332378  	& 0.341709		& 0.347747	\\ \hline
5  &  0.140301      & 0.323293	& 0.331121 		& 0.336233	\\ \hline
6   & 0.172643      & 0.314019	& 0.32075		& 0.325177	\\ \hline
7   & 0.188117      & 0.305681	& 0.311647		& 0.31559	\\ \hline
8   & 0.196376       & 0.298192	& 0.303591		& 0.307174	\\ \hline
9    & 0.200904	   & 0.291437	& 0.296399		& 0.299702	\\ \hline	
10   & 0.203298       & 0.285311	& 0.289924		& 0.293002	\\ \hline
\end{tabular}
\caption{Variation of  $\tilde{b}=\frac{b}{R}$  for  given $n=0,0.55,0.58,0.60$ with different $\lambda$}
\end{center}
\end{table}

\begin{table}
\begin{center}
\begin{tabular}{|c|c|c|c|}   \hline
$\lambda$               & $k=0.60 $	& $k=0.62$	& $k=0.63$	     \\ \hline
1   &  0.472227   & 0.497719		& 0.509691			\\ \hline
2.5 &  0.423942	 & 0.436794  	& 0.442986			\\ \hline
3  &  0.410826      & 0.422278	& 0.427808 			\\ \hline
4.5   & 0.38013      & 0.389138	& 0.393505			\\ \hline
5.6   & 0.363177      & 0.371156	& 0.375029			\\ \hline
6.1   & 0.356535       & 0.364154	& 0.367855			\\ \hline
7.5    & 0.340542	   & 0.347378	& 0.350702			\\ \hline	
8.3	& 0.332752	    & 0.339243	& 0.342401			\\ \hline
9.5	& 0.322458	    & 0.328523	& 0.331477			\\ \hline
10   & 0.318578       & 0.324491	& 0.327371			\\ \hline
\end{tabular}
\caption{Variation of  reduced size $\tilde{b}=\frac{b}{R}$  with $\lambda$ for different $k$ }
\end{center}
\end{table}

 The reduced size of a star ($\tilde{b}=\frac{b}{R}$) is tabulated for different  $n$ and $\lambda$ values in table-1. It is evident that for a given $\lambda$ if one increases $n$ the reduced size of a star increases. On the other hand for a  isotropic star as $\lambda$ increases for a given $n$ the reduced size increases but in the case of an anisotropic star the reduced size decreases in this case as one increases $\lambda$.  In table-2 reduced  size of a star is tabulated for different  $k$ and $\lambda$ values. It is evident that for a given $\lambda$ as we increase $k$ the reduced size increases. However for a given  $k$ on increasing  $\lambda$ the reduced size  of the compact object decreases.

 \section{Physical Analysis}

  For a given  mass of a compact star,   it is possible to  estimate the corresponding radius in terms of the geometric parameter $R$. To obtain stellar models we consider compact objects with observed mass \cite{tab} which  determines the radius of the  star  for different values of  $R$ with given set of values of $n$, $A$, $k$ and $\lambda$. It is known that the radius of a neutron star is $\leq (11 - 14) \;$ km. \cite{slb}, therefore, to obtain a viable stellar model for compact object the  upper bound of the size is  fixed accordingly. In the next section we consider three  stars whose masses\cite{dd,li,tab} are known from observations to explore suitability of the solutions considered here.

 {\it Model 1 } : For X-ray pulsar Her X-1 \cite{tab,dd,sdr}  characterized by  mass $m = 1.47 \; M_{\odot}$, where $M_{\odot}$  = the solar mass we obtain  a stellar configuration  with radius $r_o = 8.31106 $ km., for $R=8.169$ km. The compactness of the star in this case is  $ u=\frac{m}{r_o} = 0.30$.  The ratio of density at the boundary to that at the centre for the star is 0.128 which is satisfied for the  parameters
 $\lambda = 1.9999$,  $k=0.641$, $A=2$ and $n=0.697$. It is found that compactness factor $u=0.2$ accommodates a star of radius $r_o= 11.925 km.$  However,  stellar models with different size and compactness factor with the above mass permitted here are tabulated in  Table- 3. It is also observed that  as the compactness factor increases size of the star decreases. It is evident from the second column of Table-4 that   increase in  $\lambda$ value which is related to geometry lead to a decrease in the density profile of the compact object.

\begin{table}
\begin{center}
\begin{tabular}{|c|c|c|}   \hline
$\frac{m}{b}$    	&$R $	in km.	            & Radius ($r_o$ in km.)     \\ \hline
0.3       &8.169          &  8.311  					\\ \hline
0.28      & 8.574           &  8.828	   		  \\ \hline
0.26     & 9.048            & 9.424    				\\ \hline
0.25     & 9.317           & 9.757    				\\ \hline
0.20    & 11.096           & 11.925            \\ \hline             			
\end{tabular}
\caption{Variation of size of a star  with $\frac{m}{b}$  for   $k=0.641$, $n=0.697$, $\lambda=1.9999$  and $A=2$.}
\end{center}
\end{table}

\begin{table}[ht!]
\begin{center}
\begin{tabular}{|c|c|c|c|}   \hline
             & $\frac{\rho(b)}{\rho(0)}$ &$\frac{\rho(b)}{\rho(0)}$   &$\frac{\rho(b)}{\rho(0)}$       \\ \hline
$\lambda$   & $ n= 0.697$ & $ n= 0.60$,  & $ n= 0.50$ \\
            &   $k=0.641$  &   $k=0.63$     &   $k=0.52$  \\       \hline
1  & 0.449 & 0.508 & 0.633 \\ \hline
1.1  & 0.447 & 0.505 & 0.619 \\ \hline
1.2  & 0.444 & 0.502 & 0.607 \\ \hline
1.3  & 0.444 & 0.498 & 0.597 \\ \hline
1.4   & 0.436 & 0.494 & 0.589 \\ \hline
1.5  & 0.432 & 0.490 & 0.580 \\ \hline
1.7  & 0.429 & 0.475 & 0.565 \\ \hline
1.9999 & 0.409 & 0.466 & 0.545 \\ \hline
\end{tabular}
\caption{Density profile  $\frac{\rho(b)}{\rho(0)}$    of compact objects.}
\end{center}
\end{table}

 {\it Model 2} : For X-ray pulsar  J1518+4904 \cite{tab,dd,sdr} characterized by  mass $m = 0.72 \; M_{\odot}$, where $M_{\odot}$  = the solar mass it is noted  that it permits a star with radius $r_o = 4.071 $ km., for $R=8.169$ km. The compactness of the star in this case is  $ u=\frac{m}{r_o} = 0.30$.  The ratio of its  density at the boundary to that at the centre  is 0.142 which is obtained for values of the parameters $\lambda = 1.1$,  $k=0.641$, $A=2$ and $n=0.60$.   It is noted that a star of radius $r_o=12.332$ km. results with same mass having lower compactness factor $u= 0.09$. It is evident from Table-5 that in this case also as the compactness increases radius of the star decreases. The variation of density profile with $\lambda$ is tabulated  in the 3rd column of Table -4. It is found that the density profile decreases as $\lambda$ increases.

\begin{table}[ht!]
\begin{center}
\begin{tabular}{|c|c|c|}   \hline
$\frac{m}{b}$    	&$R $	in km.	            & Radius($r_o$ in km.)     \\ \hline
0.3       &8.169          &  4.071 					\\ \hline
0.28      & 8.574           &  4.324	   		  \\ \hline
0.26     & 9.048            & 4.616    				\\ \hline
0.24     & 9.317           & 4.956    				\\ \hline
0.22  & 11.096           & 5.358           \\ \hline              			
\end{tabular}
\caption{Variation of size of a star  with $\frac{m}{b}$  for   $k=0.63$, $n=0.60$, $\lambda=1.1$  and $A=2$.}
\end{center}
\end{table}

 {\it Model 3} : In this case we consider a compact object B1855+09(g)
  \cite{tab,dd,sdr}  characterized by  mass $m = 1.6 \; M_{\odot}$, where $M_{\odot}$  = the solar mass, it is noted that its radius is $r_o = 9.047$ km., for $R=8.169$ km. with compactness factor  $ u=\frac{m}{r_o} = 0.30$.  The ratio of density at the boundary to that at the centre for the star is 0.187 which is found  for the values of the parameters $\lambda = 1$,  $k=0.52$, $A=2$ and $n=0.50$.  It is noted that a star of compactness factor $u=0.22$ accommodates a star with radius $r_o= 11.907$ km. For the same mass considered here it is possible to obtain a class of stellar models  with different size and  compactness which are tabulated in Table-6. We note that size of the star decreases with the increase in compactness.  The variation  of density profile with  $\lambda$  is displayed in 4th column of Table- 4. It is evident the density profile  decreases as $\lambda$  increases.

\begin{table}[ht!]
\begin{center}
\begin{tabular}{|c|c|c|}   \hline
$\frac{m}{b}$    	&$R $	in km	            & Radius($r_o$ in km.)     \\ \hline
0.3       &8.169          &  9.047 					\\ \hline
0.28      & 8.574           &  9.609	   		  \\ \hline
0.26     & 9.048            & 9.818    				\\ \hline
0.24     & 9.317           & 11.013    				\\ \hline
0.22  & 11.096           & 11.907           \\ \hline              			
\end{tabular}
\caption{Variation of size of a star  with $\frac{m}{b}$  for   $k=0.63$, $n=0.60$, $\lambda=1.1$  and $A=2$.}
\end{center}
\end{table}

\begin{table}[ht!]
\begin{center}
\begin{tabular}{|c|c|}  \hline
\multicolumn{1}{|c|}{Star with mass}   & \multicolumn{1}{|c|}{Radial pressure}  \\  \hline
HER X-1              \\ \hline

  1.47$M_{\odot}$               &(i) $p_r=1.207\rho-8.477$        \\
                                  & (ii) $p_r=0.130\rho^2-1.032\rho+0.980$\\ \hline
 J1518+4904                             \\ \hline

   0.72$M_{\odot}$               &(i) $p_r=1.041\rho-7.607 $          \\
                                      &(ii)  $p_r=0.104\rho^2-0.794\rho+0.350$ \\ \hline
B1855+09(g)                              \\ \hline
1.6$M_{\odot}$                           &(i) $p_r=0.602\rho-5.316 $          \\
										&(ii) $p_r=0.043\rho^2-0.252\rho-1.151$ \\ \hline
\end{tabular}
\caption{ Variation of radial pressure with density for different stellar models.}
\end{center}
\end{table}

\begin{figure}
\begin{center}
\includegraphics{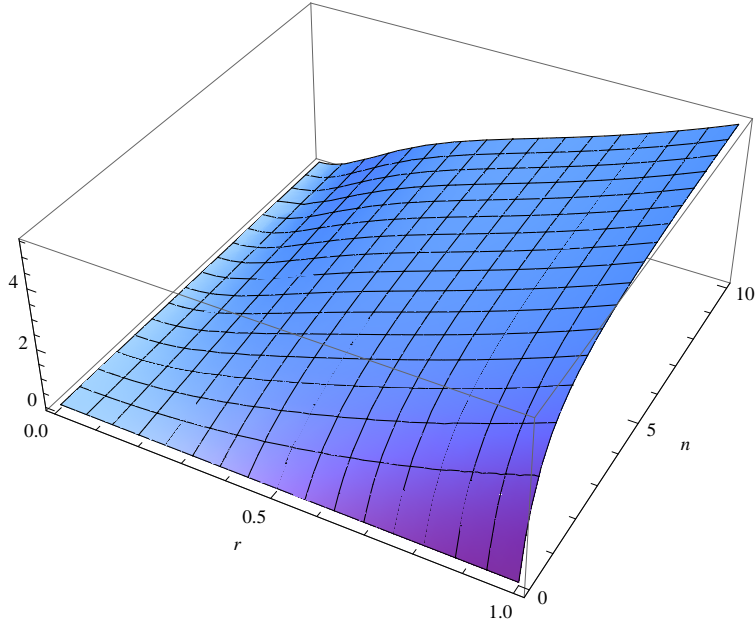}
\caption{Plot of $\Delta$ with positive $n$ and radial distance   with $\lambda=2$ and $k=0.4$. }
\end{center}
\end{figure}


\section{Discussion}

In this paper, we present a class of new general relativistic solutions for a class of compact stars which are in hydrostatic equilibrium considering an   anisotropic interior fluid distribution. The radial pressure and the tangential pressure are  different, variations of the pressures are determined. As the EOS of the fluid inside a neutron star is not known  so we adopt here numerical technique to determine a suitable EOS of the matter content inside the star for a given space-time geometry. The interior space-time geometry considered here is  characterized by five geometrical parameters namely,  $\lambda$, $R$, $k$, $A$  and $n$ which are used to obtain different stellar models.  For $n=0$, the relativistic solution reduces to that considered in  by \cite{ps} and \cite{rb}.   The permitted  values of the unknown  parameters are determined from the  following conditions : (a) metric matching at the boundary, (b) $\frac{dp}{d\rho}<1$ , (c)  pressure at the boundary is zero i.e., $p=0$ and (d) the positivity of density.

We note the following:
(i) In figs. (1) and (2),  the radial variation of pressure and density are plotted for different  $k$ for a given set of values of $\lambda$, $A$, $n$  and $k$. The radial pressure  increase with an increase in $k$ but the density is found to decrease.   The central density of the compact object increases if $k$ decreases. (ii) In fig. (3),  variation of radial pressure inside the star is plotted for different  $n$. We note that pressure decreases as   $n$  increases, however, density does not change.
(iii) In figs. (4) and (5), radial variation of density and pressure are plotted for different  $\lambda$. We note that both the pressure  and the density  increases with an increase in $\lambda$. The   central  density is found to  increase with an increase in $\lambda$ in this case. The radial variation of pressure for different $\lambda$ is shown in fig. (5). It is  noted that the radial pressure near the boundary decreases sharply for higher values of $\lambda$.

(iv) It is evident from  figs. (8) and (9) that SEC is obeyed inside the stars for the configurations considered in the two cases. In  figs. (10)-(12)  we obtain an interesting result where SEC is violated. The size of the region near the centre is further  increases with an increase in the value of one of the parameters,  $n$, $k$ and $\lambda$  keeping the other parameters unchanged. Thus the solution obtained here may be useful to construct  a core-envelope  model of a compact star    which will be discussed elsewhere.
 (v) In fig. (13), the  radial variation of anisotropy inside the star for different $n$ values are plotted.  The increase in  value of $n$ is related to  increase in anisotropy of the fluid pressure.
 (vi) For a given $\lambda$ as we increase $n$ the reduced size  of  star increases. However for $n=0$ the size of a star increases with an increase in $\lambda$ which is tabulated in  Table-1. It is noted that  for  non-zero values of $n$ the size of the star however found to decrease.

 (vii) For a given $\lambda$ the size of the star increases as $k$ increases, but for a given $k$ the size of the star decreases as $\lambda$ increases which is shown in Table-2.
 (viii) Considering observed masses  of  the compact objects namely, HER X-1, J1518+4904 and B1855+09(g) we explore the interior of the star. A class of compact stellar models with anisotropic pressure distribution are permitted with the new solution discussed here. In the models  stars of different  compactness factor which are shown in Tables-(3),(5)and (6) for different geometric parameters. The density profile of the models are also tabulated in Tables-(4).  The density profile inside the star is found to decrease as   $\lambda$ increases.
 (ix) We obtain functional relation of the radial pressure with the density for the models considered here which is presented in  Table-(7). It is noted that a viable stellar model may be obtained here with a polynomial EOS. In the table we have displayed linear and a quadratic EoS only, it may be mentioned here that similar EoS are considered recently in \cite{mps} and \cite{cdp} to obtain relativistic stellar models. We note that though a stellar configuration in our case permits a linear EoS, it does not accommodate a star satisfying MIT bag model \cite{cdp}. It is also noted  that the stellar  models obtained here allows neutron stars with mass less than $2 M_{\odot}$ for an anisotropic fluid distribution.  The observed maximum mass of a neutron star is $2 M_{\odot}$, therefore the stellar models obtained here may be relevant for compact objects with nuclear density. A physically realistic stellar model up to radius $(11 \sim 14)$ km.  may be permitted here with the relativistic solutions accommodating less compactness  \cite{slb}.
 (x) We plot radial variation of the anisotropy measurement in pressure {\it i.e.}, $\Delta$ in fig. (14)  with $n$. It is evident from the 3D plot  that  $\Delta \rightarrow 0$ when $n  \rightarrow 0$ which leads to isotropic pressure case. For $n > 0$, the difference in tangential pressure to radial pressure initially increases which however attains a constant value for large $n$.

\vspace{1.0 cm}

{ \it Acknowledgement :}

BCP would like to acknowledge  TWAS-UNESCO for supporting a visit to Institute of Theoretical Physics, Chinese Academy of Sciences, Beijing where the work is completed. BCP would like to thank  University Grants Commission, New Delhi for financial support (Grant no. F.42-783/2013(SR)). RD is also thankful to UGC, New Delhi and Physics Department, North Bengal University for providing research facilities. The authors would like to thank the referee for constructive  suggestion.

\pagebreak

\end{document}